\documentclass[letterpaper,twoside,12pt]{article}
\usepackage{amsmath,amssymb}
\usepackage{latexsym}
\usepackage{verbatim}
\usepackage{floatfig,epsfig}
\usepackage{calc}
\usepackage{amstext,amsfonts}
\usepackage{fancyheadings}
\usepackage{pstricks,pst-node}
\newlength{\treescale}
\newlength{\treeunit}
\newcommand{\papertitle}{%
The Hopf Algebra of Renormalization, \\[3mm]
Normal Coordinates and \\[3mm]
Kontsevich Deformation Quantization%
}
\newcommand{\headtitle}{%
The Hopf Algebra of Renormalization..%
}
\newcommand{\paperauthor}{%
M.{} Rosenbaum and J.{} D.{} Vergara%
} 
\newlength{\enviropost}
\newcommand{\be}{\begin{equation}}
\newcommand{\ee}{\end{equation}}
\newcommand{\ble}[1]{\begin{equation} \label{#1}}
\newcommand{\bae}{\begin{eqnarray}}
\newcommand{\eae}{\end{eqnarray}}
\newcommand{\fle}[2]%
{\vspace{1.5ex} \be \label{#1}
\mbox{%
\setlength{\fboxsep}{3ex}%
\framebox{$\dss #2 $}} \ee}

\newcommand{\Tone}[2]{
  \begin{picture}(27,36)(0,0)
  \put(2,5){\psfig{figure=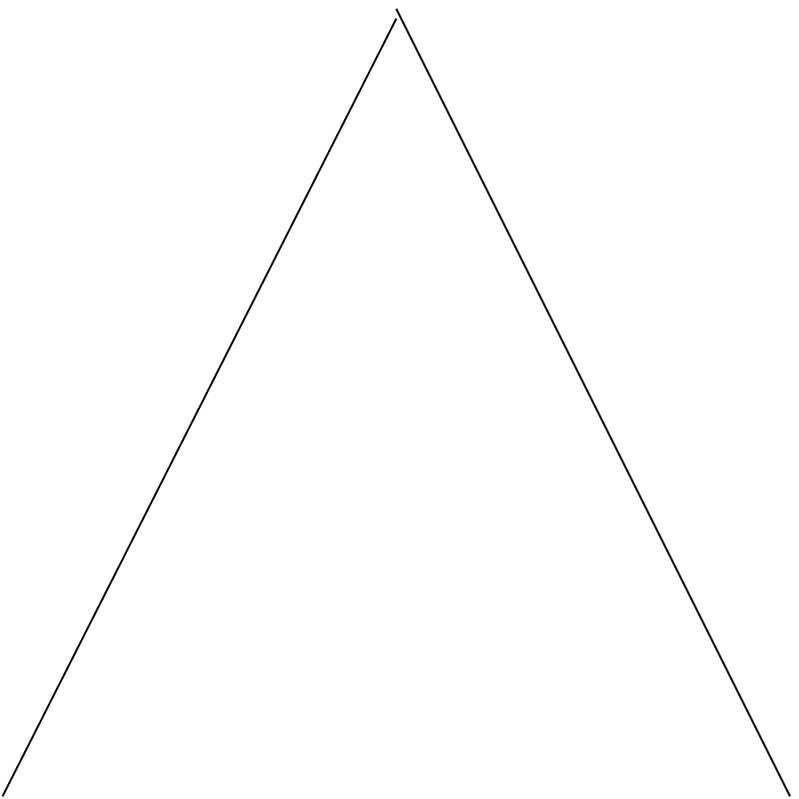,width=17pt,angle=0}}
  \put(17,-3){$\scriptstyle   #1$}
  \put(0,-3){$\scriptstyle   #2$}
   \end{picture}}
\newcommand{\Ttwol}[3]{
  \begin{picture}(27,37)(0,0)
  \put(2,5){\psfig{figure=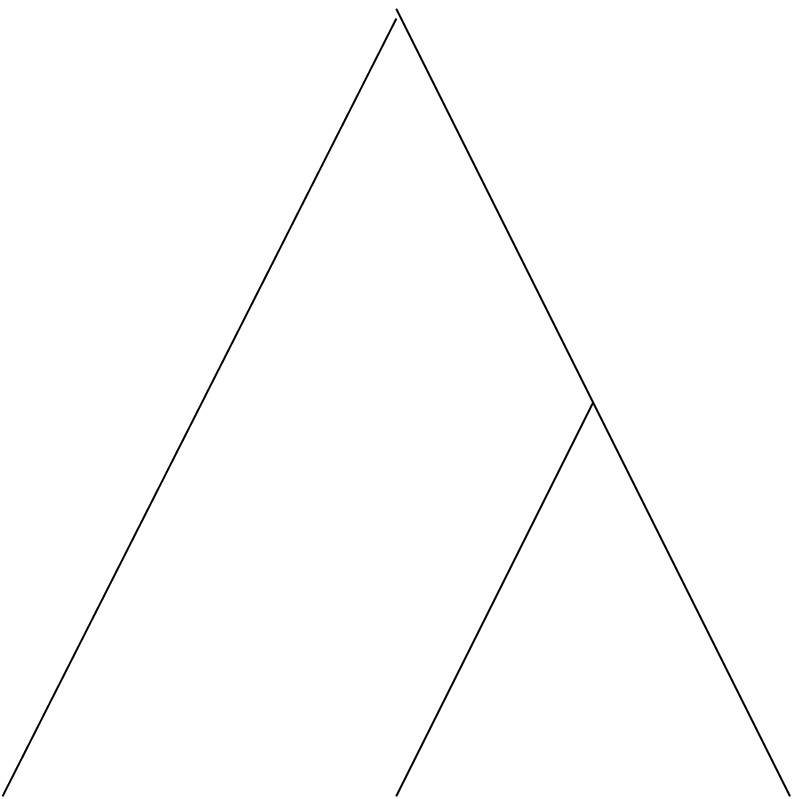,width=20pt,angle=0}}
  \put(0,-3){$\scriptstyle   #1$}
  \put(10,-3){$\scriptstyle  #2$}
  \put(20,-3){$\scriptstyle #3$}
\end{picture}}
\newcommand{\Ttiwol}[7]{
  \begin{picture}(30,43)(0,0)
  \put(2,5){\psfig{figure=2l.eps,width=22pt,angle=0}}
  \put(0,-3){$\scriptstyle   #1$}
  \put(10,-3){$\scriptstyle  #2$}
  \put(20,-3){$\scriptstyle #3$}
  \put(-1,19){$\scriptstyle   #4$}
  \put(17,19){$\scriptstyle   #5$}
  \put(4,6){$\scriptstyle   #6$}
  \put(23,6){$\scriptstyle   #7$}
\end{picture}}
\newcommand{\Ttbbnn}[6]{
  \begin{picture}(34,51)(0,0)
  \put(2,2){\psfig{figure=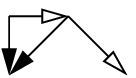,width=31pt,height=29pt,angle=0}}
  \put(0,-3){$\scriptstyle   #1$}
  \put(26,-3){$\scriptstyle  #2$}
  \put(-6,18){$\scriptstyle #3$}
  \put(4,18){$\scriptstyle   #4$}
  \put(23,18){$\scriptstyle   #5$}
  \put(6,31){$\scriptstyle   #6$}
\end{picture}}
\newcommand{\Ttvvww}[4]{
  \begin{picture}(28,45)(0,0)
  \put(2,5){\psfig{figure=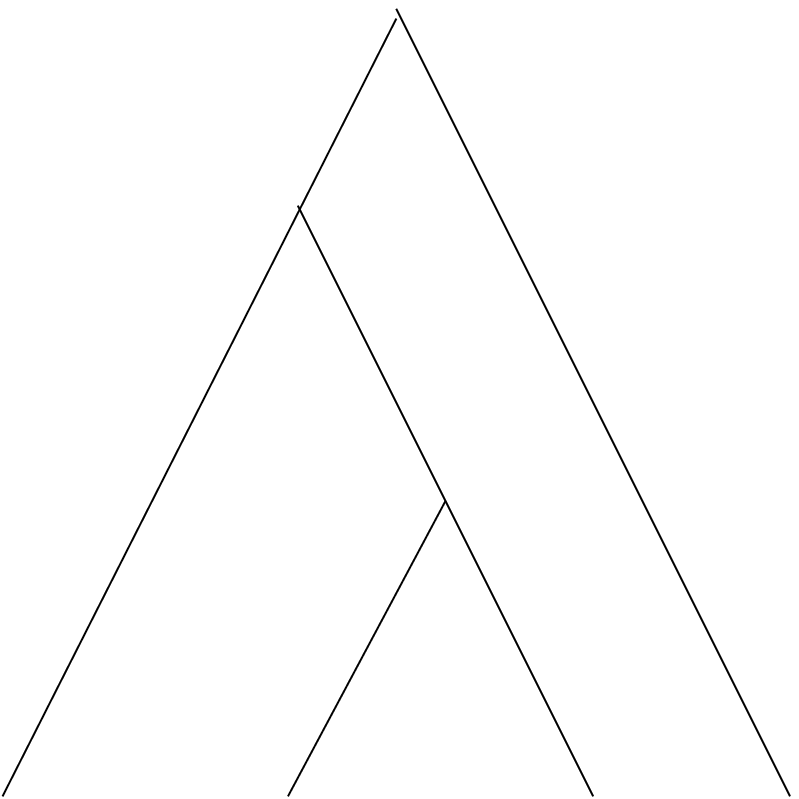,width=23pt,angle=0}}
  \put(0,-2){$\scriptscriptstyle  #1$}
  \put(7,-2){$\scriptscriptstyle   #2$}
  \put(15,-2){$\scriptscriptstyle  #3$}
  \put(22,-2){$\scriptscriptstyle #4$}
   \end{picture}}
\newcommand{\Ttivvww}[9]{
  \begin{picture}(30,52)(0,0)
  \put(2,5){\psfig{figure=3l.eps,width=28pt,angle=0}}
  \put(0,-1){$\scriptscriptstyle  #1$}
  \put(8,-1){$\scriptscriptstyle   #2$}
  \put(17,-1){$\scriptscriptstyle  #3$}
  \put(26,-1){$\scriptscriptstyle #4$}
  \put(4,28){$\scriptscriptstyle #5$}
  \put(18,28){$\scriptscriptstyle #6$}
  \put(-2,16){$\scriptscriptstyle #7$}
  \put(8,16){$\scriptscriptstyle #8$}
  \put(4,5){$\scriptscriptstyle #9$}
  \end{picture}}
\newcommand{\Tbbbnnn}[8]{
  \begin{picture}(34,51)(0,0)
  \put(2,2){\psfig{figure=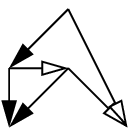,width=31pt,height=32pt,angle=0}}
  \put(0,-3){$\scriptstyle   #1$}
  \put(24,-3){$\scriptstyle  #2$}
  \put(4,28){$\scriptscriptstyle #3$}
  \put(20,27){$\scriptscriptstyle   #4$}
  \put(10,21){$\scriptscriptstyle   #5$}
  \put(-6,11){$\scriptscriptstyle   #6$}
  \put(10,6){$\scriptscriptstyle #7$}
  \put(19,5){$\scriptscriptstyle #8$}
\end{picture}}
\newcommand{\Tibbbnnn}[8]{
  \begin{picture}(34,51)(0,0)
  \put(2,2){\psfig{figure=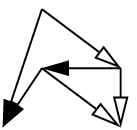,width=31pt,height=32pt,angle=0}}
  \put(0,-3){$\scriptstyle   #1$}
  \put(24,-3){$\scriptstyle  #2$}
  \put(1,28){$\scriptscriptstyle #3$}
  \put(20,27){$\scriptscriptstyle   #4$}
  \put(13,21){$\scriptscriptstyle   #5$}
  \put(9,11){$\scriptscriptstyle   #6$}
  \put(15,5){$\scriptscriptstyle #7$}
  \put(31,11){$\scriptscriptstyle #8$}
\end{picture}}

\newcommand{\Tivvww}[4]{
  \begin{picture}(28,45)(0,0)
  \put(2,5){\psfig{figure=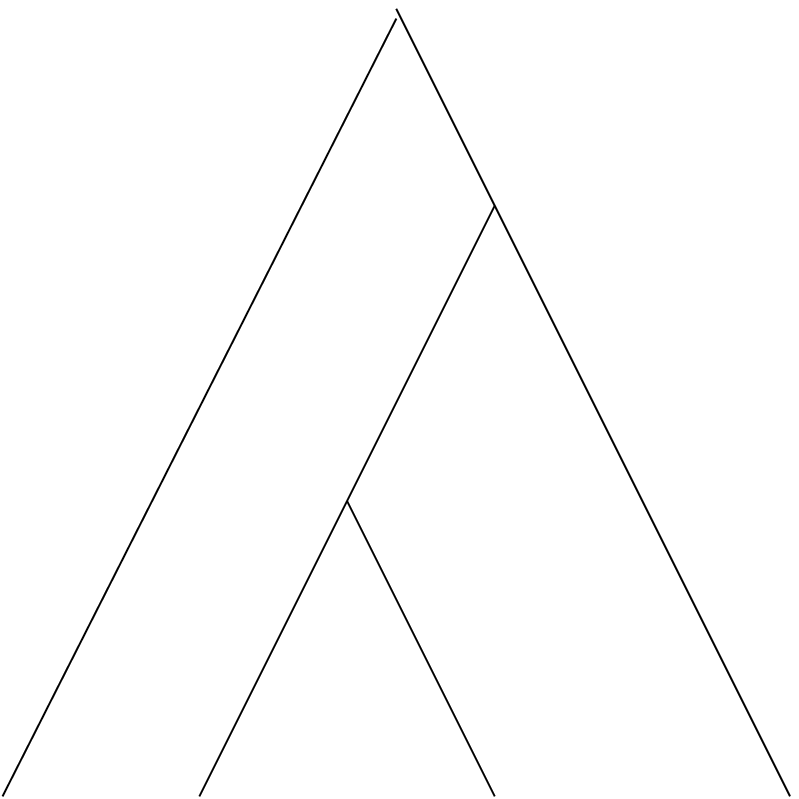,width=23pt,angle=0}}
  \put(0,-1){$\scriptscriptstyle  #1$}
  \put(7,-1){$\scriptscriptstyle   #2$}
  \put(13,-1){$\scriptscriptstyle  #3$}
  \put(21,-1){$\scriptscriptstyle #4$}
   \end{picture}}
\newcommand{\Tvvww}[9]{
  \begin{picture}(30,52)(0,0)
  \put(2,5){\psfig{figure=3il.eps,width=28pt,angle=0}}
  \put(0,-1){$\scriptscriptstyle  #1$}
  \put(7,-1){$\scriptscriptstyle   #2$}
  \put(15,-1){$\scriptscriptstyle  #3$}
  \put(24,-1){$\scriptscriptstyle #4$}
  \put(3,27){$\scriptscriptstyle #5$}
  \put(19,27){$\scriptscriptstyle #6$}
  \put(15,16){$\scriptscriptstyle #7$}
  \put(25,16){$\scriptscriptstyle #8$}
  \put(10,5){$\scriptscriptstyle #9$}
  \end{picture}}

\newcommand{\Tloop}[9]{
  \begin{picture}(90,95)(0,0)
  \put(0,15){\psfig{figure=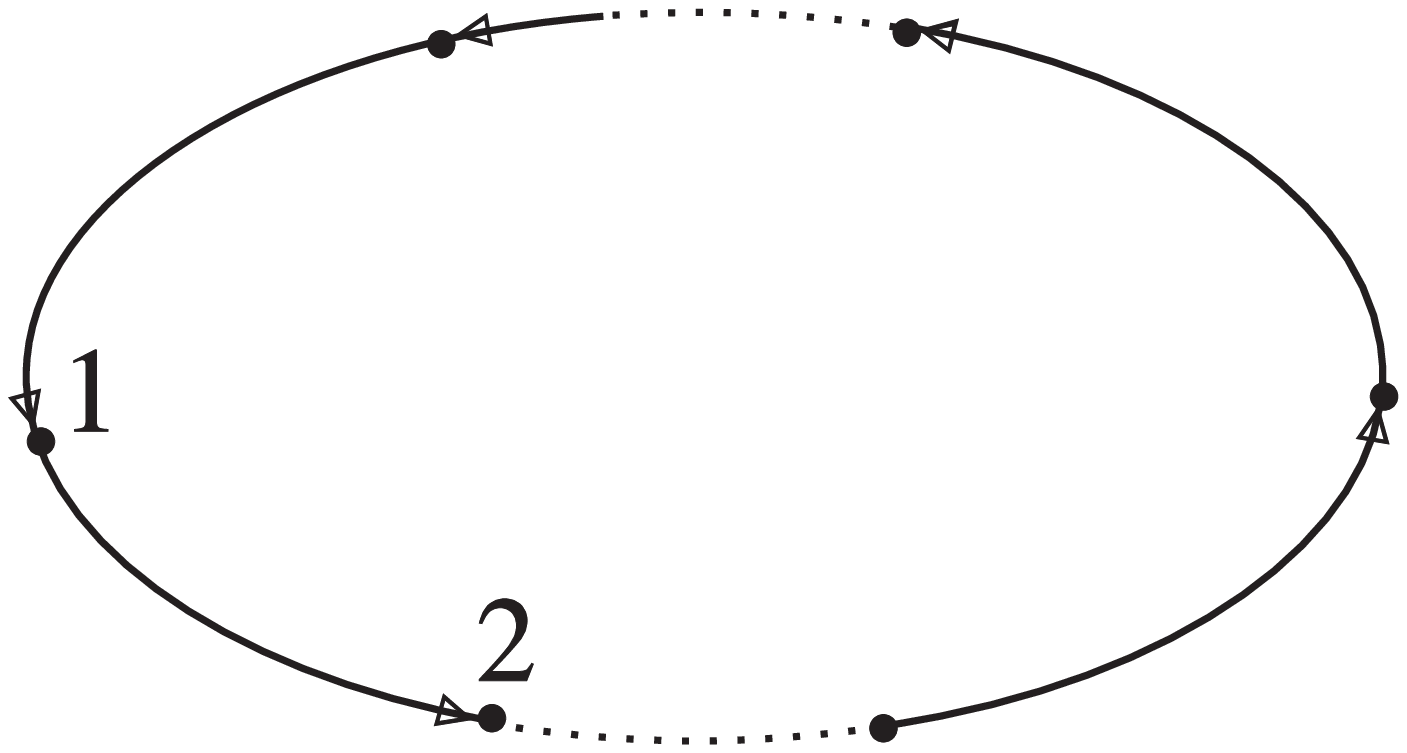,width=90pt,angle=0}}
  \put(26,68){$\scriptscriptstyle  #1$}
  \put(25,55){$\scriptstyle   #2$}
  \put(-11,49){$\scriptscriptstyle  #3$}
  \put(-36,36){$\scriptscriptstyle #4$}
  \put(3,21){$\scriptscriptstyle #5$}
  \put(20,7){$\scriptscriptstyle #6$}
  \put(50,18){$\scriptscriptstyle #7$}
  \put(89,36){$\scriptscriptstyle #8$}
  \put(62,62){$\scriptscriptstyle #9$}
  \end{picture}}

\newcommand{\Ttwo}[3]{
  \begin{picture}(27,37)(0,0)
  \put(2,5){\psfig{figure=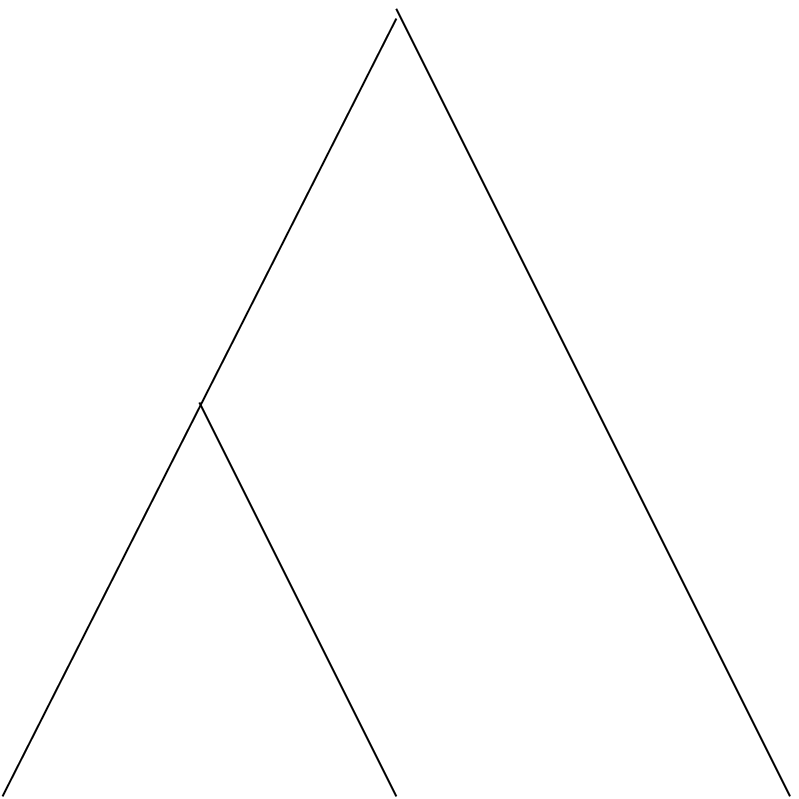,width=20pt,angle=0}}
  \put(0,-3){$\scriptstyle   #1$}
  \put(10,-3){$\scriptstyle  #2$}
  \put(20,-3){$\scriptstyle  #3$}
   \end{picture}}

\newcommand{\Tgammab}[3]{
  \begin{picture}(30,38)(0,0)
  \put(2,5){\psfig{figure=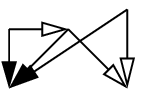,width=25pt,height=26pt,angle=0}}
  \put(0,-3){$\scriptstyle   #1$}
  \put(20,-3){$\scriptstyle  #2$}
   \end{picture}}
\newcommand{\Ttiwo}[7]{
  \begin{picture}(32,43)(0,0)
  \put(6,3){\psfig{figure=2il.eps,width=22pt,angle=0}}
  \put(3,-3){$\scriptstyle   #1$}
  \put(13,-3){$\scriptstyle  #2$}
  \put(23,-3){$\scriptstyle #3$}
  \put(3,19){$\scriptstyle   #4$}
  \put(21,19){$\scriptstyle   #5$}
  \put(-3,6){$\scriptstyle   #6$}
  \put(16,6){$\scriptstyle   #7$}
\end{picture}}
\newcommand{\Ttnnbb}[6]{
  \begin{picture}(34,51)(0,0)
  \put(2,2){\psfig{figure=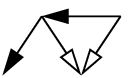,width=31pt,height=29pt,angle=0}}
  \put(0,-3){$\scriptstyle   #1$}
  \put(20,-3){$\scriptstyle  #2$}
  \put(-2,18){$\scriptstyle #3$}
  \put(16,18){$\scriptstyle   #4$}
  \put(30,18){$\scriptstyle   #5$}
  \put(20,31){$\scriptstyle   #6$}
\end{picture}}
\newcommand{\Albc}[3]{
  \begin{picture}(27,37)(0,0)
  \put(2,0){\psfig{figure=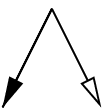,width=20pt,angle=0}}
  \put(-1,11){$\scriptstyle   #1$}
  \put(10,22){$\scriptstyle  #2$}
  \put(20,11){$\scriptstyle #3$}
\end{picture}}
\newcommand{\Albcd}[5]{
  \begin{picture}(27,37)(0,0)
  \put(2,0){\psfig{figure=nb.eps,width=20pt,angle=0}}
  \put(-2,-4){$\scriptstyle   #1$}
  \put(20,-4){$\scriptstyle  #2$}
  \put(-1,11){$\scriptstyle   #3$}
  \put(10,22){$\scriptstyle  #4$}
  \put(20,11){$\scriptstyle #5$}
\end{picture}}
\newcommand{\Alvw}[2]{
  \begin{picture}(17,17)(0,0)
  \put(0,0){\psfig{figure=nb.eps,width=14pt,angle=0}}
  \put(-2,-4){$\scriptstyle   #1$}
  \put(10,-4){$\scriptstyle  #2$}
\end{picture}}
\newcommand{\Alvwt}[2]{
  \begin{picture}(27,30)(0,0)
  \put(0,0){\psfig{figure=nb.eps,width=25pt,angle=0}}
  \put(-2,-4){$\scriptstyle   #1$}
  \put(20,-4){$\scriptstyle  #2$}
\end{picture}}

\newtheorem{Def}{Definition}

\newtheorem{exatitle}{Example}
{\noindent \textsc{Proof}:\ }%
{\hfill $\blacksquare$  \vspace{\enviropost} \\}
{\begin{exatitle} \label{#2} #1 \end{exatitle}}%
{\hfill $\Box$ \vspace{\enviropost} \\}
{\vspace{3mm} \noindent {\bf Aside:} \small}%
{\hfill \rule[-3mm]{0mm}{0mm}$\Diamond$\\}

\newcommand{\dss}{\displaystyle}

\newcommand{\eg}{\hbox{\it e.g.{}}}

\newcommand{\slte}{\setlength{\treescale}{1.8\treeunit}}

\begin{document}
\normalsize \initfloatingfigs
\begin{titlepage}
\vspace*{-1cm}
\begin{flushright}
\textsf{ICN-UNAM-04/08}
\\
\mbox{}
\\
\textsf{April 28, 2004}
\\[3cm]
\end{flushright}
\renewcommand{\thefootnote}{\fnsymbol{footnote}}
\begin{LARGE}
\bfseries{\sffamily \papertitle}
\end{LARGE}

\noindent \rule{\textwidth}{.6mm}

\vspace*{1.6cm}

\noindent
\begin{large}%
\textsf{\bfseries%
\paperauthor }
\end{large}


\phantom{XX}
\begin{minipage}{.8\textwidth}
\begin{it}
\noindent Instituto de Ciencias Nucleares \\
Universidad Nacional Aut\'onoma de M\'exico\\
Apdo. Postal 70-543, 04510 M\'exico, D.F., MEXICO \\
\end{it}
\texttt{mrosen,vergara@nuclecu.unam.mx \phantom{X}}
\end{minipage}
\\

\vspace*{3cm}
\noindent \textsc{\large Abstract: }Using normal coordinates in
 a Poincar\'e-Birkhoff-Witt basis for the Hopf algebra of renormalization
 in perturbative quantum field theory, we investigate the relation
between the twisted antipode axiom in that formalism, the Birkhoff algebraic
decomposition and the universal formula of Kontsevich for
quantum deformation.

\noindent PACS: 02.40Gh, 11.10.Gh, 03.70.+k, 03.65.Fd

\noindent Keywords: Renormalization, Star products, Quantum field
theory.
\end{titlepage}
\setcounter{footnote}{1}
\renewcommand{\thefootnote}{\arabic{footnote}}
\setcounter{page}{2}
\noindent \rule{\textwidth}{.5mm}

\tableofcontents

\noindent \rule{\textwidth}{.5mm}
\newsavebox{\Treeoo}
\newsavebox{\Treeto}
\newsavebox{\Treetho}
\newsavebox{\Treetht}
\newsavebox{\Treefo}
\newsavebox{\Treeft}
\newsavebox{\Treefth}
\newsavebox{\Treeff}
\newsavebox{\Treefio}
\newsavebox{\Treefit}
\newsavebox{\Treefith}
\newsavebox{\Treefif}
\newsavebox{\Treefifi}
\newsavebox{\Treefis}
\newsavebox{\Treefise}
\newsavebox{\Treefie}
\newsavebox{\Treefin}
\sbox{\Treeoo}{{%
\begin{pspicture}(2mm,-1mm)(8mm,10mm)
\psset{xunit=1cm,yunit=1cm} \psdots[dotscale=3](.5,0)
\end{pspicture}}%
}
\newcommand{\Too}[1][0]{%
\raisebox{#1\totalheight}%
  {\resizebox{!}{\treescale}{\usebox{\Treeoo}}}}
\sbox{\Treeto}{{%
\begin{pspicture}(2mm,-1mm)(8mm,10mm)
\psset{xunit=1cm,yunit=1cm}
\psline[linewidth=.3mm]{*-*}%
(.5,1)(.5,0) \psdots[dotscale=3](.5,1)(.5,0)
\end{pspicture}}%
}
\newcommand{\Tto}[1][0]{%
\raisebox{#1\totalheight}%
  {\resizebox{!}{\treescale}{\usebox{\Treeto}}}}
\sbox{\Treetho}{{%
\begin{pspicture}(2mm,-1mm)(8mm,2cm)
\psset{xunit=1cm,yunit=1cm}
\psline[linewidth=.3mm]{*-*}%
(.5,2)(.5,0) \psdots[dotscale=3](.5,2)(.5,1)(.5,0)
\end{pspicture}}%
}
\newcommand{\Ttho}[1][0]{%
\raisebox{#1\totalheight}%
  {\resizebox{!}{2\treescale}{\usebox{\Treetho}}}}
\sbox{\Treetht}{{%
\begin{pspicture}(-3mm,-1mm)(13mm,10mm)
\psset{xunit=1cm,yunit=1cm}
\psline[linewidth=.3mm]{*-*}%
(.5,1)(0,0)
\psline[linewidth=.3mm]{*-*}%
(.5,1)(1,0) \psdots[dotscale=3](.5,1)(0,0)(1,0)
\end{pspicture}}%
}
\newcommand{\Ttht}[1][0]{%
\raisebox{#1\totalheight}%
  {\resizebox{!}{\treescale}{\usebox{\Treetht}}}}
\sbox{\Treefo}{{%
\begin{pspicture}(2mm,-1mm)(8mm,30mm)
\psset{xunit=1cm,yunit=1cm}
\psline[linewidth=.3mm]{*-*}%
(.5,3)(.5,0) \psdots[dotscale=3](.5,3)(.5,2)(.5,1)(.5,0)
\end{pspicture}}%
}
\newcommand{\Tfo}[1][0]{%
\raisebox{#1\totalheight}%
  {\resizebox{!}{3\treescale}{\usebox{\Treefo}}}}
\sbox{\Treefth}{{%
\begin{pspicture}(2mm,-1mm)(13mm,20mm)
\psset{xunit=1cm,yunit=1cm}
\psline[linewidth=.3mm]{*-*}%
(.5,2)(.5,0)
\psline[linewidth=.3mm]{*-*}%
(.5,2)(1,1) \psdots[dotscale=3](.5,2)(.5,1)(.5,0)(1,1)
\end{pspicture}}%
}
\newcommand{\Tfth}[1][0]{%
\raisebox{#1\totalheight}%
  {\resizebox{!}{2\treescale}{\usebox{\Treefth}}}}
\sbox{\Treeft}{{%
\begin{pspicture}(-3mm,-1mm)(13mm,20mm)
\psset{xunit=1cm,yunit=1cm}
\psline[linewidth=.3mm]{*-*}%
(.5,2)(.5,1)
\psline[linewidth=.3mm]{*-*}%
(.5,1)(0,0)
\psline[linewidth=.3mm]{*-*}%
(.5,1)(1,0) \psdots[dotscale=3](.5,2)(.5,1)(0,0)(1,0)
\end{pspicture}}%
}
\newcommand{\Tft}[1][0]{%
\raisebox{#1\totalheight}%
  {\resizebox{!}{2\treescale}{\usebox{\Treeft}}}}
\sbox{\Treeff}{{%
\begin{pspicture}(-3mm,-1mm)(13mm,10mm)
\psset{xunit=1cm,yunit=1cm}
\psline[linewidth=.3mm]{*-*}%
(.5,1)(0,0)
\psline[linewidth=.3mm]{*-*}%
(.5,1)(.5,0)
\psline[linewidth=.3mm]{*-*}%
(.5,1)(1,0) \psdots[dotscale=3](.5,1)(0,0)(.5,0)(1,0)
\end{pspicture}}%
}
\newcommand{\Tff}[1][0]{%
\raisebox{#1\totalheight}%
  {\resizebox{!}{\treescale}{\usebox{\Treeff}}}}
\sbox{\Treefio}{{%
\begin{pspicture}(2mm,-1mm)(8mm,40mm)
\psset{xunit=1cm,yunit=1cm}
\psline[linewidth=.3mm]{*-*}%
(.5,4)(.5,0) \psdots[dotscale=3](.5,4)(.5,3)(.5,2)(.5,1)(.5,0)
\end{pspicture}}%
}
\newcommand{\Tfio}[1][0]{%
\raisebox{#1\totalheight}%
  {\resizebox{!}{\treescale}{\usebox{\Treefio}}}}
\section{Introduction}
\label{Intro}

A regular feature in frontier physics and mathematics has been the
passage from commutative to non-commutative
structures\cite{Nekra}, and deformation quantization has been a
major factor in this trend ( for a nice review of the genesis,
developments and major metamorphoses in this field we refer the
reader to the paper by Dito and Sternheimer in \cite{Dito}). A
medular contribution to the metamorphosis of deformation
quantization has been the work of Kontsevich \cite{Kont:97} and
the proof therein of his Formality Theorem, which allowed to
establish the existence of star
associative products on general Poisson differentiable manifolds.\\
The most obvious example of the relevance in physics of
deformation quantization is the Moyal product, based on a constant
Poisson structure, and which exhibits the passage from classical
to quantum mechanics as a deformation of the pointwise product of
smooth functions on ${\mathbb R}^d$ in the direction of the
Poisson product. It is well known that the Moyal deformation
operator is the exponential of a
bi-differential.\\
Next in the order of complexity are the linear Poisson structures
for which the paradigm is the Lie-Poisson bracket first introduced
by S. Lie himself, and latter rediscovered by F. Berezin and A.
Kirillov. In fact, the analysis of a linear Poisson structure on
${\mathbb R}^d$ is equivalent to considering the vector space dual
to a Lie algebra with the Poisson structure induced by the Lie
bracket of the algebra. For this linear Poisson structure there
are at least two canonical quantization deformations known: The
universal formula of Kontsevich (equivalent to the Duflo star
product\cite{Duflo}) and the one arising from the classical
Baker-Campbell-Hausdorff (BCH) formula. The BCH product corresponds
to the Gutt product \cite{Gutt} obtained from the product of elements in the
universal enveloping algebra via the symmetrization operator,
while in the Kontsevich construction (which for clarity purposes we review in
Section 5) each term in the product
corresponds to a graph, associated to a bi-differential operator,
and all graphs have a weight defined by the integration of a
$2n-$form, where $n$ is the set of edges of the graph.\\
The relation between these two quantizations has been considered
recently by various authors
\cite{Arnal,Kathotia,Dito1,Shoikhet,Polyak}. It was shown by
Kathotia \cite{Kathotia} that the BCH formula is exactly that part
of the Kontsevich formula consisting of all the admissible
L-graphs without wheels and that the two quantizations are totally
equivalent for the case of
nilpotent Lie algebras. \\

Based on the relation of the above mentioned quantization deformations, the
Hopf algebraic formulation of renormalization in perturbative quantum field
theory (pQFT), first discovered by Kreimer \cite{Kre:98} and further developed by
Connes and Kreimer \cite{Con.Kre:98,Con.Kre:00}, together with our introduction of
the concept of normal coordinates in the Hopf algebra of renormalization \cite{Chry},
we shall show
here that the Forest Formula for renormalization in pQFT and Birkhoff's
algebraic decomposition in that context, can be interpreted,
for any renormalizable field theory, as a Kontsevich star-product deformation
in the direction of the Lie-Poisson product, and where the Kontsevich bi-differential
operator is an exponential of a sum of admissible prime L-graphs.\\

Indeed, the Connes-Kreimer formalism involves two Hopf algebras: The (commutative,
but not co-commutative) Hopf algebra ${\mathcal H}_R$ generated by representatives of
decorated rooted trees, and the (non-commutative, co-commutative)
Hopf algebra ${Char\mathcal H}_{R}$  of the group of characters in duality with
${\mathcal H}_R$ and
isomorphic to the universal enveloping algebra $U(\mathfrak L)$ of a Lie
algebra $\mathfrak L$. But, on the one hand, the deformation quantization of the
universal envelope of a Lie algebra corresponds to group multiplication via the BCH
formula and, on the other hand,  the use of normal coordinates in the construction of
a basis for  ${\mathcal H}_R$ introduces a group product and BCH formula in the
definition of the coproduct for the normal coordinates. Consequently, since
such a coproduct appears in the twisted antipode axiom for renormalization within
the Connes-Kreimer and normal coordinates formalism ({\it cf.} Eq.(\ref{2.7f}) below),
it becomes reasonable to expect a relation between renormalization in
pQFT and the bi-differential symplectic operator of Kontsevich for quantum
deformations in the case of a linear Poisson structure. The key point in
this observation is that although the Lie algebra  $\mathfrak L$ is not
nilpotent all the wheels are null. \\

\section{The Hopf algebra of Renormalization, Characters,  Infinitesimal
Characters and Normal Coordinates}

As a basis for our discussion we shall make use of the normal
coordinates for the Hopf algebra of renormalization, which we
previously introduced in \cite{Chry}. So in order to fix notation
and make our presentation as self-contained as possible we begin
by reviewing some of the relevant results in that paper which we
shall be making use of here.\\

One-particle irreducible superficially divergent Feynman diagrams
in pQFT can be represented by decorated rooted trees (or sums of
them for the case of overlapping divergences) which are finite,
connected graphs without loops where every vertex has one incoming
edge except for the root that has only outgoing edges. The
decorations of the vertices are primitive diagrams (divergent but
without subdivergences) \cite{Con.Kre:98}. Let ${\mathcal
H}_{R}(m, {\bf 1}=e1, \Delta, \epsilon, S)$ denote the graded
commutative (but not co-commutative) Hopf algebra, over a field
$\mathbb K$ of characteristic zero, generated by the rooted trees.
By the Milnor-Moore theorem, there is a co-commutative Hopf
algebra ${\bf G}={Char\mathcal H}_{R}$ in duality with ${\mathcal
H}_{R}$, isomorphic to the universal enveloping algebra
$U(\mathfrak L)$ where $\mathfrak L = \partial Char{\mathcal
H}_{R}$ is a Lie algebra. ${\bf G}$ is the group of characters of
${\mathcal H}_{R}$ (algebra morphism under the convolution product
$\langle \eta\ast\lambda, T^{A} \rangle
:=\langle\eta\otimes\lambda, \Delta T^{A} \rangle,\;\;\eta,
\lambda \in {\bf G}$, with $T^{A}$ a representative of an
isomorphism class of rooted trees \eg \slte \ble{Tree1} T^{A} \in
\{ \Too , \Tto  , \Ttho  , \Ttht , \Tfo , \Tft , \Tfth , \Tff ,
\dots \}). \ee


Let $Z_{A}$ denote the infinitesimal generators of $\mathfrak L$
indexed by rooted trees and defined by
\begin{eqnarray}
\langle Z_{A}, T^{B} \rangle &=& \delta_{A}^{B},\\\label{1.1}
\langle Z_{A}, T^{B} T^{C} \rangle &=& \langle Z_{A}, T^{B} \rangle
\epsilon(T^{C}) +\epsilon(T^{B})\langle Z_{A}, T^{C} \rangle.
\label{1.2}
\end{eqnarray}
Since the coproduct in $U(\mathfrak L)$ is dual to the product in
${\mathcal H}_{R}$ we have
\begin{equation}
\begin{split}
\langle Z_{A}\ast Z_{B} ,T^{C} \rangle= \langle Z_{A}\otimes Z_{B}, \Delta T^{C} \rangle= \sum_{T} n_{T^{A} T^{B}}^{T} \langle Z_T,T^{C} \rangle,\\
\;\; Z_{A}, Z_{B}\in \partial Char{\mathcal H}_{R},\label{1.3}
\end{split}
\end{equation}
which defines a pre-Lie algebra on $\partial Char{\mathcal
H}_{R}$, and the Lie bracket
\begin{equation}
\begin{split}
[Z_{A}, Z_{B}]:=Z_{A}\ast Z_{B}- Z_{B}\ast Z_{A}= \sum_{T} (n_{T^{A} T^{B}}^{T} - n_{T^{B} T^{A}}^{T}) Z_T\\
\equiv \sum_{T} f_{T^{A} T^{B}}^{T} Z_T, \label{1.4}
\end{split}
\end{equation}
where $n_{T^{A} T^{B}}^{T}$ is the number of simple cuts on $T$ that produce
the sub-trees $T^{A}$ and $T^{B}$,
 with $T^{B}$ containing the root of $T$. The last equality in (\ref{1.4})
 defines the structure
 constants $f_{T^{A} T^{B}}^{T}$ of $\mathfrak L$.\\

Now, if \eg \slte \ble{Tree2} \{f_{i} \}= \{ {\bf 1},  \Too , \Tto
, \Too\Too , \Ttho  , \Ttht , \Too \Tto , \Too\Too\Too , \dots \}
\ee is a given Poincar\'e-Birkhoff-Witt basis for ${\mathcal
H}_{R}$, we can obtain a dual basis $\{e^{i}\}$ for the enveloping
algebra $U(\mathfrak L)$ by adjoining to the above $Z$'s
polynomials in them (via the convolution product given by
(\ref{1.3})) with $\langle e^i, f_{j}\rangle =\delta^{i}_{j}$. For
the basis dual to (\ref{Tree2}) we get (for the case of vertices
with the same decoration)
\begin{equation}
\begin{split}
\{e^{i}\}=\{{\bf 1}, Z_{\Too}, Z_{\Tto}, \frac{1}{2}(Z_{\Too}\ast
Z_{\Too}- Z_{\Tto}), Z_{\Ttho}, Z_{\Ttht},\\ -Z_{\Ttho}+
Z_{\Ttht}-\frac{1}{2} Z_{\Too}\ast Z_{\Tto}+\frac{3}{2}
Z_{\Tto}\ast Z_{\Too},
 \frac{1}{6}Z_{\Too}\ast Z_{\Too}\ast Z_{\Too}, \dots\}. \label{1.6}
\end{split}
\end{equation}
Clearly the calculation of the elements of the basis (\ref{1.6})
becomes increasingly more complicated with increasing degree
(number of vertices in the trees). We can change however the basis
for $U(\mathfrak L)$ to the simpler one
\begin{equation}
\{e^{i^{\prime}}\}=\{{\bf 1}, Z_{A}, Z_{A}\ast Z_{B}, \dots
\}=\{{\bf 1}, Z_{\Too},\dots, Z_{\Too}\ast Z_{\Too}, \dots,
\}.\label{1.7}
\end{equation}
In order to construct its dual, let ${\tilde \psi}^{A}$ be new
coordinates centered at the origin and indexed by rooted trees.
Choose then a new linear basis with the following ordering
\begin{equation}
 \{f_{i^{\prime}} \}= \{ {\bf 1}, {\tilde \psi}^{A}, {\tilde \psi}^{A}{\tilde \psi}^{B},\dots \}.\label{1.8}
\end{equation}
Since $\{e^{i^{\prime}}\}$ and $\{f_{i^{\prime}} \}$ are by
construction dual to each other, the canonical tensor
\begin{equation}
{\bf C}=\sum_{i} f_{i^{\prime}}\otimes e^{i^{\prime}}=
e_{\ast}^{{\tilde \psi}^{A}\otimes Z_{A}},\label{1.9}
\end{equation}
where the $\ast$-exponential, defined by
\begin{equation}
e_{\ast}^{x} = \sum_{i=0}^{\infty}
\frac{1}{i!}\underbrace{x\ast\dots \ast x}_{i \;\;\text
{factors}}, \;\;\; x\in {\mathcal U}_{1}, \label{1.9b}
\end{equation}

acts as an identity on $T^{A}$, {\it ie.}
\begin{equation}
\langle e_{\ast}^{{\tilde \psi}^{B}\otimes Z_{B}}, {\text
id}\otimes T^{A} \rangle= T^{A}.\label{1.10}
\end{equation}
Hence
\begin{equation}
T^{A}= \sum_{m=0}^{\infty} \frac{1}{m!} {\tilde
\psi}^{T^{B_1}}\dots {\tilde \psi}^{T^{B_m}} \langle
Z_{B_1}\ast\dots\ast Z_{B_m}, T^{A} \rangle. \label{1.11}
\end{equation}
Moreover, since
\begin{equation}
\langle Z_{B_1}\ast\dots\ast Z_{B_m}, T^{A} \rangle =\langle
Z_{B_1}\otimes\dots\otimes Z_{B_m}, \Delta^{m-1} (T^{A})
\rangle, \label{1.12}
\end{equation}
where the higher powers of the convolution product are defined iteratively by
\begin{equation}
\Delta^{(0)}=\text {id}, \;\;\;\; \Delta^{(m)}:=(\text {id} \otimes \Delta^{(m-1)})\circ \Delta,\label{cop}
\end{equation}
we have that
\begin{equation}
T^{A}= {\tilde \psi}^{A}+ \sum_{m=2}^{\infty}
\frac{1}{m!}{\tilde \psi}^{B_1}\dots {\tilde
\psi}^{B_m} n_{B_{1}\dots B_{m}}^{A}, \label{1.13}
\end{equation}
with
\begin{equation}
n_{B_{1}\dots B_{m}}^{A}= n_{B_{1} R_{1}}^{A} n_{B_{2}
R_{2}}^{R_{1}} \dots n_{B_{m-1} B_{m}}^{R_{m-2}}, \label{1.14}
\end{equation}
and summation over repeated indices understood throughout.\\
The relation between the above two sets of generators can be
expressed concisely by
\begin{equation}
e_{\ast}^{{\tilde \psi}^{A}\otimes Z_{A}}= T^{B}\otimes
Z_{B}.\label{1.15}
\end{equation}

It is not difficult to see from (\ref{1.11}) that the matrix
relating the Poincar\'e -Birkhoff-Witt bases $\{ f_i\}$ and $\{
f_{i^{\prime}}\}$ is upper triangular with units along the
diagonal, and therefore invertible ({\it cf.} \cite{Chry}).
Consequently the normal coordinates ${\tilde \psi}^{A}$ are
polynomials in terms of rooted trees with the linear part equal to
$T^{A}$.
\\
The Hopf structure for ${\mathcal H}_{R}$ in terms of normal
coordinates is derived by recalling the standard property of ${\bf
C}$:
\begin{equation}
(\Delta\otimes \text {id}) e_{\ast}^{{\tilde \psi}^{A}\otimes
Z_{A}} =e_{\ast}^{{\Delta(\tilde \psi}^{A})\otimes
Z_{A}} =e_{\ast}^{{\tilde \psi}^{B}\otimes {\bf 1}
\otimes Z_{B}}\ast e_{\ast}^{{\bf 1}\otimes {\tilde
\psi}^{C}\otimes Z_{C}},\label{1.16}
\end{equation}
and applying the Baker-Campbell-Hausdorff (BCH) formula to the
group product on the right side of (\ref{1.16}). Thus the
coproduct $\Delta ({\tilde \psi}^{A})$ is given by the coefficient
of $Z_{A}$ in the resulting Hausdorff series in the exponent.
Explicitly
\begin{equation}
\Delta ({\tilde \psi}^{A})= {\tilde \psi}^{A}\otimes {\bf
1} + {\bf 1}\otimes {\tilde \psi}^{A}+\frac{1}{2}f_{B_1
B_2}^{A}{\tilde \psi}^{B_1}\otimes {\tilde
\psi}^{B_2}+\dots . \label{1.17}
\end{equation}
Because the counit on $T^{A}$ vanishes ($\epsilon (T^{A})=0,\;\; T^{A} \neq {\bf 1}$), it clearly follows from (\ref{1.13}) that the same is true for the ${\tilde \psi}^{A}$.\\

Recalling now that the exponential map is a bijection from $\partial
Char{\mathcal H}_{R} \rightarrow Char{\mathcal H}_{R}$, and that
the inverse of a character $\xi = e^{\alpha^{A}Z_{A}}$ is given
by  $\xi^{-1} = \xi\circ S$, we can derive an expression for
the action of the antipode $S$ on the normal coordinates. Thus
\begin{equation}
\langle e_{\ast}^{-\alpha^{A}Z_{A}}, {\tilde \psi}^{A}\rangle
=\langle \xi^{-1}, {\tilde \psi}^{A}\rangle = \langle
\xi\circ S, {\tilde \psi}^{A})\rangle =\langle \xi,
S({\tilde \psi}^{A})\rangle \label{1.18}
\end{equation}
and, since by the definition of normal coordinates
\begin{equation}
\langle \xi, {\tilde \psi}^{A}\rangle=\alpha^{A},
\label{1.19}
\end{equation}
it readily follows that
\begin{equation}
S({\tilde \psi}^{A})=- {\tilde \psi}^{A}.\label{1.20}
\end{equation}
\section{Renormalization in terms of normal coordinates}
\label{Hartd}
The standard approach to renormalization in pQFT is to first regularize the theory by mapping the expressions corresponding to the Feynman diagrams onto analytic functions on the Riemann sphere ${\bf PC}^{1}$. In the Hopf algebra approach to renormalization this implies considering the homomorphisms from the algebra of decorated rooted trees, ${\mathcal H}_{R}$, to the unital ${\mathbb C}$-algebra ${\mathcal A}=\{f\in \text {Holom} ({\mathbb C} -0)\}$ with 0 a pole of finite order.\\
Let $\varphi, \varphi^{\prime}\in \text {Hom}_{\mathbb {C}-\text
{alg.}} ({\mathcal H}_{R}, {\mathcal A})$ be two such linear maps
$\varphi,\varphi^{\prime}:\mathbb {C} ({\mathcal
H}_{R})\rightarrow {\mathcal A}$. Multiplication in this unital
${\mathbb C}$-algebra of homomorphisms is defined by the
convolution product
\begin{equation}
(\varphi\ast\varphi^{\prime})({T^{A}})=m_{\mathcal
A}(\varphi\otimes\varphi^{\prime})(\Delta {T^{A}}),\label{2.1}
\end{equation}
which correspondingly implies for the normal coordinates
\begin{equation}
(\varphi\ast\varphi^{\prime})({\tilde \psi}^{A})=m_{\mathcal
A}(\varphi\otimes\varphi^{\prime})(\Delta {\tilde
\psi}^{A}),\label{2.2}
\end{equation}
with the coproducts on the bi-algebras of  rooted trees and normal
coordinates defined, respectively, by admissible cuts of branches
in rooted trees (for details {\it cf. e.g.} \cite{Con.Kre:00}) and
by equation (\ref{1.17}) above. Note that by setting
$\varphi({\tilde \psi}^{A})(z):= \psi^{A}\in {\mathcal A},\;\;
\varphi(T^{A})(z):=\phi^{A} \in {\mathcal A}$, for $z\in \mathbb {C}
-\{0\}$, we can make contact with the notation in \cite{Chry}.\\

In the Hopf algebra of renormalization the equivalent to the
Forest Formula is the twisted antipode axiom:
\begin{eqnarray}
\phi^{A}_{R}& =&m_{\mathcal A}\circ (S_{\mathcal R} \otimes \text {id})(\varphi\otimes\varphi)(\Delta {T^{A}}), \nonumber\\
\psi^{A}_{R}& =&m_{\mathcal A}\circ (S_{\mathcal R} \otimes \text
{id})(\varphi\otimes\varphi)(\Delta {\tilde \psi}^{A}).
\label{2.3}
\end{eqnarray}
Here $\phi^{A}_{R}$ and $\psi^{A}_{R}\;$ stand for the
renormalized $\phi^{A}$ and $\psi^{A}$, respectively, while
$\mathcal R$ is the linear map ${\mathcal R}:\text {Hom}_{\mathbb
{C}-\text {alg.}} ({\mathcal H}_{R}, {\mathcal A})\rightarrow
\text {Hom}_{\mathbb {C}-\text {alg.}} ({\mathcal H}_{R},
{\mathcal A})$, by $\varphi \mapsto {\mathcal R}(\varphi):= R\circ
\varphi : {\mathcal H}_{R}\rightarrow R({\mathcal A})$, and $R$ is
a Rota-Baxter projection operator, chosen to give the pole part of
its argument (mass independent renormalization scheme),
 which satisfies the multiplicative constraints
\begin{equation}
R(ab)+R(a)R(b)=R \left ( R(a)b+a R(b)\right ),  \;\;\; a,b\in
{\mathcal A}.\label{2.4}
\end{equation}
This makes ${\mathcal A}$ a Rota-Baxter algebra of weight one. The multiplicative twisted antipode $S_{\mathcal R}$ is
defined recursively by
\begin{eqnarray}
S_{\mathcal R}(\phi^{A}) &=& -R\left [\phi^{A}+ m_{\mathcal A}\circ (S_{\mathcal R}\otimes \text {id})(\varphi\otimes\varphi){\tilde \Delta}(T^{A})\right ]\nonumber\\
S_{\mathcal R}(\psi^{A}) &=& -R\left [\psi^{A}+ m_{\mathcal A}
\circ (S_{\mathcal R}\otimes \text {id}))\varphi\otimes\varphi)\tilde
{\Delta}(\tilde{\psi}^{A})\right ].\label{2.5}
\end{eqnarray}
In the above equations the symbol $\tilde {\Delta}$ is used to denote the
coproduct with the primitive part omitted. \\

Note that the target space of the counterterm map $S_{\mathcal R}:
{\mathcal A}\to{\mathcal A}_{-}$ is
$${\mathcal A}_{-} =\{\text {polynomials in $z^{-1}$ without constant term}\},$$
{\it ie.} the principal part of the Laurent series for the $\phi^{A}$ or $\psi^{A}$,
respectively.

Let us now apply the operator $\;(m_{\mathcal A} \otimes \text {id})\circ (S_{\mathcal R}\otimes \text {id}\otimes \text {id})\circ (\varphi\otimes\varphi\otimes \text {id})\circ(\Delta\otimes \text {id})\;$ to equation (\ref{1.15}) and make use of (\ref{1.16}) and the commutativity of ${\mathcal A}$. We thus get
\begin{equation}
\begin{split}
(\phi^{B})_{R} \otimes Z_{B}= e_{\ast}^{S_{\mathcal R}(\psi^{A})\otimes Z_{A}}\ast e_{\ast}^{\psi^{C}\otimes Z_{C}}\\
= e_{\ast}^{[S_{\mathcal R}(\psi^{A}) + \psi^{A}+ \frac{1}{2} f_{B_{1} B_{2}}^{A} S_{\mathcal R}(\psi^{B_{1}}) \psi^{B_{2}} + \dots ]\otimes Z_{A}} = e_{\ast}^{(\psi^{A})_{R} \otimes Z_{A}},\label{2.7f}
\end{split}
\end{equation}
where the term in the exponential in the equality before the last is the Hausdorff
series for the BCH product.\\

\section{Renormalization and Bi-differential Operators}

  Write (\ref{2.7f}) as
\begin{equation}
e_{\ast}^{V}\ast e_{\ast}^{W}=e^{H(V, W)},\label{2.10f}
\end{equation}
where $V=S_{\mathcal R}(\psi^{A})\otimes Z_{A}$ ,
$W=\psi^{C}\otimes Z_{C}$. The Hausdorff series $H(V, W)$ has the properties\\

(i)$H=\sum_{n=1}^{\infty} H_{n}$, with $H_{n}$ the homogeneous
part of $H$ of degree $n$ in $W$ given by \cite{Reut}:
\begin{eqnarray}
H_n &=& \frac{1}{n!}\left( H_1 \frac{\partial}{\partial V}\right)^n (V),\label{zpoisson1}\\
H_1 &=& W+\frac{1}{2}[V, W]+ \sum_{k=1}\frac{B_{2k}}{(2k)!}\;\text {ad}(V)^{2k} (W)\label{zpoisson2},
\end{eqnarray}
where $B_{2k}$ are the Bernoulli numbers. Note however, that due to redundancies stemming from skew-symmetry and the Jacobi identities  the resulting formula for the Hausdorff series obtained by this procedure, as well as for all other known presentations, has a non-minimal character;\\

(ii) $H(V, H(W,Z))=H(H(V, W),Z)$ for $V,W,Z \in {\mathcal A}\otimes {\mathfrak L} \}$
(associativity);\\

(iii) $H(V, -V)=0$, $H(V, 0)=H(0, V)=V$.\\

Consider next the free Lie algebra on the
two generators $V,\; W$. By property (i) above, we can rewrite
$H$ as $H=\sum_{n=1}^{\infty} H^{(n)}$, where $ H^{(n)}$ are now
finite linear combinations of multi-commutators formed from $n$-letter
words. Thus using the natural
graphical encoding provided by Hall trees for Lie algebras
\cite{Reut}, we have that each $H^{(n)}$ may be represented by a
finite linear combination of Hall trees. Specifically:
\begin{equation}
H^{(2)}(V, W)=\frac{1}{2}[V, W]=\frac{1}{2} \Tone{W}{V}
\:\;;\nonumber
\end{equation}
\begin{equation}
H^{(3)}(V, W)=\frac{1}{12}([V,[V, W]]+[[V,W], W])=\frac{1}{12}\left
(\Ttwol{V}{V}{W}+ \Ttwo{V}{W}{W}\right );\nonumber
\end{equation}
\begin{equation}
H^{(4)}(V, W)=\frac{1}{48}\left ([V, [[V,W], W]]+[[V,[V, W]],W]
\right)=\frac{1}{48}
\left(\Tivvww{V}{V}{W}{W}+\Ttvvww{V}{V}{W}{W}\right);\nonumber
\end{equation}
and so on to higher orders.\\

If we now let ${\mathfrak L}^{\ast}$ denote the Lie algebra dual
to ${\mathfrak L}$, it is well
known that ${\mathfrak L}^{\ast}$ can be equipped with a linear Poisson structure,
induced by the Lie bracket on ${\mathfrak L}$ \cite{Takh}, given by
$f_{B_{1} B_{2}}^{A} Z^{\ast}_{A}\; \partial_{B_{1}}\partial_{B_{2}}$,
where the $Z^{\ast}_{A}$ are coordinates on  ${\mathfrak L}^{\ast}$ and the
partials are taken relative to this coordinates.\\
 Thus, starting with the BCH formula (\ref{2.10f}), we can  introduce a
 $\star_1$-product for the algebra of exponential functions $e^{Z^{\ast}_{A}\psi^{A}}$
 and rewrite (\ref{2.7f}) in the form:
\begin{equation}
\begin{split}
e^{(\psi^{A})_{R} Z^{\ast}_{A}}=
e^{S_{\mathcal R} (\psi^{B})Z^{\ast}_{B}}{\star}_{1} e^{\psi^{C}Z^{\ast}_{C}}:=\hspace{3in}\\
e^{S_{\mathcal R} (\psi^{B})Z^{\ast}_{B}}\;\hat{D}\left (Z^{\ast}, f, ({\overleftarrow{\partial}}_{B_{1}}, {\overleftarrow{\partial}}_{B_{2}}\dots), ({\overrightarrow{\partial}}_{C_{1}}, {\overrightarrow{\partial}}_{C_{2}}\dots)\right ) \;e^{\psi^{C}Z^{\ast}_{C}}, \label{2.8f}
\end{split}
\end{equation}
where $\hat{D}$ is the bi-differential operator:
\begin{equation}
\begin{split}
\hat{D}\left (Z^{\ast}, f, ({\overleftarrow{\partial}}_{B_{1}},
{\overleftarrow{\partial}}_{B_{2}}\dots), ({\overrightarrow{\partial}}_{C_{1}},
{\overrightarrow{\partial}}_{C_{2}}\dots)\right )=\hspace{3in}\\
e^{\{\left [\frac{1}{2}
{\overleftarrow{\partial}}_{B_{1}} f_{B_{1}  C_{1}}^{A_{1}}{\overrightarrow{\partial}}_{C_{1}}+
\frac{1}{12}(
{\overleftarrow{\partial}}_{B_{1}} {\overleftarrow{\partial}}_{B_{2}}
f_{B_{1}  C_{1}}^{C_{2}} f_{B_{2} C_{2}}^{A_{1}}{\overrightarrow{\partial}}_{C_{1}}+ {\overleftarrow{\partial}}_{B_{1}}f_{B_{1}  C_{1}}^{A_{2}}
 f_{A_{2} C_{2}}^{A_{1}}
 {\overrightarrow{\partial}}_{C_{1}} {\overrightarrow{\partial}}_{C_{2}})
 +\dots \}\right ]Z^{\ast}_{A_{1}}}. \label{2.9f}\\
\end{split}
\end{equation}

One can immediately conclude that the above $\star_1$-product is associative
by computing the right side term in (\ref{2.8f}) and comparing with the BCH formula
for the product of two group elements; the results are the same and, since the
Hausdorff series is associative (by property (ii) above), so is the $\star_1$-product.\\
Moreover, the term on the right of the first equality in (\ref{2.8f})
is a sum of $\star_1$-products of polynomials on ${\mathfrak L}^{*}$ of the form
$p_m \star_1 q_{n}, \;\;n,m=0,1,2,\dots$ which satisfy the properties:\\
\begin{itemize}
  \item (A 1) For any two polynomials $p_m$ and $q_n$ of orders $m$ and $n$,
  respectively, the $\star_1$-product satisfies
\begin{equation}
p_n \star_1 q_m =p_n q_m + r_{n+m-1},\label{polynom}
\end{equation}
where $p_n q_m$ is the pointwise product of $p_m$ and $q_n$ and $r_{n+m-1}$ is a
polynomial of degree $n+m-1$.\\
This property follows from the fact that the bi-differential
operator
$\hat D$ in (\ref{2.9f}) reduces the total degree by one, except for the pointwise
product resulting from the zero-th order in the expansion of the exponential in
(\ref{2.9f}).\\

  \item (A 2) From (A 1) and the associativity of the $\star_1$-product it
  follows that for any two $p_m$ and $q_n$ the product $p_m \star_1 q_{n}$ is
  determined by knowing $(V)^{n}\star_1 W,\;\; n\in {\mathbb N}$, where without risk of confusion we now use the same symbols $V, W$ to express  $V=S_{\mathcal R}(\psi^{A})Z^{*}_{A}$, and
$W=\psi^{C}Z^{*}_{C}$, in terms of coordinates of ${\mathfrak L}^{*}$.\\
The proof of this property is based on induction and a polarization identity. For
further details of the proof we refer the reader to Lemma 2.1.1 in \cite{Kathotia}.\\

\end{itemize}

Let us now consider the transition from Hall trees, encoding the Lie
algebra in $H(V,W)$,
to graphs for bi-differential operators colored by a linear
Poisson structure. This is discussed and illustrated extensively in
\cite{Kathotia}, so here we shall only summarize the procedure which
consists essentially in the following three steps:

\begin{itemize}
  \item Coloring the wedges in the Hall tress with the Poisson
  structure.
  \item Identifying the basic ordered wedge  $\Albc{B_i}{\alpha^{B_i C_j}} {C_j}$
  with the bi-differential operator \break
$\overleftarrow\partial_{B_i}(\alpha^{B_i
C_j})\overrightarrow\partial_{C_j}$, {\it i.e.}
\begin{equation}\label{albc}
\overleftarrow\partial_{B_i}(\alpha^{B_i
C_j})\overrightarrow\partial_{C_j} = \Albc{B_i}{\alpha^{B_i
C_j}}{C_j},
\end{equation}
where $\alpha^{B_i C_j}\equiv \frac{1}{2} f^{A_i}_{B_i C_j}
Z^{*}_{A_i} $ and where we assign the arrow $-\!\!\!-\!\!\!\rhd$
to the right action $\overrightarrow\partial_{C_j}$ and the black
arrow $-\!\!\!-\!\!\!\blacktriangleright$ to the left action
$\overleftarrow\partial_{B_i}$.
  \item Merging all the V's in the Hall trees into one point (to the left) and all
  the W's into another point (to the right).\\

 Thus for example to
  order $H^{(4)}$ in the Hausdorff series we get:
 \begin{eqnarray}
\!\!\!\! [V,W]\equiv \text {ad}_{V}(W) \Rightarrow \Tone{W}{V}
\Rightarrow \Albcd{V}{W}{B_1}{\alpha^{B_1
C_1}}{C_1}&\Leftrightarrow & e^{V}\left(\overleftarrow\partial_{B_1}
f^{A_1}_{B_1
C_1}Z^*_{A_1}\overrightarrow\partial_{C_1}\right)e^W \nonumber\\
&=& e^V \left(2 \overleftarrow\partial_{B_1}\alpha^{B_1
C_1}\overrightarrow\partial_{C_1}\right) e^{W} ;  \label{hs.1}
\end{eqnarray}

  \begin{eqnarray}
 [V,[V,W]]\equiv \text {ad}^{2}_{V}(W) \Rightarrow
\Ttwol{V}{V}{W} &\Rightarrow& \Ttiwol{V}{V}{W}{B_2}{C_2}{B_1}{C_1}
\Rightarrow \Ttbbnn{V}{W}{B_2}{B_1}{C_1}{C_2} \nonumber\\
&\Leftrightarrow& e^{V} \left(\overleftarrow\partial_{B_2}
\overleftarrow\partial_{B_1}f^{C_2}_{B_1C_1}f^{D_2}_{B_2C_2}Z^*_{D_2}
\overrightarrow\partial_{C_1}\right)e^{W}  \nonumber \\
&=& e^{V} \left((2)^2 \overleftarrow\partial_{B_2}
\overleftarrow\partial_{B_1}(\partial_{C_2}\alpha^{B_1C_1})\alpha^{B_2C_2}
\overrightarrow\partial_{C_1}\right)e^{W}\;\;; \nonumber \\
\label{hs.2}
\end{eqnarray}
  \begin{eqnarray}
[[V,W],W]\equiv \text {ad}_W (\text {ad}_{V}(W)) \Rightarrow
\Ttwo{V}{W}{W} &\Rightarrow& \Ttiwo{V}{W}{W}{B_2}{C_2}{B_1}{C_1}
\Rightarrow \Ttnnbb{V}{W}{B_1}{C_1}{C_2}{B_2}\nonumber\\
&\Leftrightarrow& e^{V}\left(\overleftarrow\partial_{B_1}
f^{B_2}_{B_1C_1}f^{D_2}_{B_2C_2}Z^*_{D_2}
\overrightarrow\partial_{C_1}\overrightarrow\partial_{C_2}\right)e^{W} \nonumber\\
&=& e^{V}\left((2)^2 \overleftarrow\partial_{B_1}
(\partial_{B_2}\alpha^{B_1C_1})\alpha^{B_2C_2}
\overrightarrow\partial_{C_1}\overrightarrow\partial_{C_2}\right)e^{W}\;\;;\nonumber\\
\label{hs.3}
\end{eqnarray}
  \begin{eqnarray}
[V,[[V,W],W]&\equiv& \text {ad}_V (\text {ad}_W (\text
{ad}_{V}(W))) \Rightarrow \Tivvww{V}{V}{W}{W} \Rightarrow
\Tvvww{V}{V}{W}{W}{B_2}{C_2}{B_3}{C_3}{B_1\ \!\! C_1} \Rightarrow
\Tibbbnnn{V}{W}{B_2}{C_2}{B_3}{B_1}{C_1}{C_3}  \nonumber\\
&\Leftrightarrow&
e^{V}\left(\overleftarrow\partial_{B_2}\overleftarrow\partial_{B_1}
f^{B_3}_{B_1C_1}f^{C_2}_{B_3C_3}f^{D_2}_{B_2C_2}Z^*_{D_2}
\overrightarrow\partial_{C_1}\overrightarrow\partial_{C_3}\right)e^{W}  \nonumber\\
&=&
e^{V}\left((2)^3 \overleftarrow\partial_{B_2}\overleftarrow\partial_{B_1}
(\partial_{B_3}\alpha^{B_1C_1})(\partial_{C_2}\alpha^{B_3C_3})\alpha^{B_2C_2}
\overrightarrow\partial_{C_1}\overrightarrow\partial_{C_3}\right)e^{W}\;\;;\nonumber\\
\label{hs.4}
\end{eqnarray}
  \begin{eqnarray}
[[V,[V,W]],W]&\equiv& \text {ad}_W (\text {ad}^{2}_{V}(W))
\Rightarrow \Ttvvww{V}{V}{W}{W} \Rightarrow
\Ttivvww{V}{V}{W}{W}{B_2}{C_2}{B_3}{C_3}{B_1\ \! \! C_1}
\Rightarrow \Tbbbnnn{V}{W}{B_2}{C_2}{C_3}{B_3}{B_1}{C_1} \nonumber
\\ &\Leftrightarrow&
e^{V}\left(\overleftarrow\partial_{B_3}\overleftarrow\partial_{B_1}
f^{C_3}_{B_1C_1}f^{D_2}_{B_2C_2}f^{B_2}_{B_3C_3}Z^*_{D_2}
\overrightarrow\partial_{C_1}\overrightarrow\partial_{C_2}\right)e^{W}  \nonumber\\
&=&
e^{V}\left((2)^3 \overleftarrow\partial_{B_3}\overleftarrow\partial_{B_1}
(\partial_{C_3}\alpha^{B_1C_1})(\partial_{B_2}\alpha^{B_3C_3})\alpha^{B_2C_2}
\overrightarrow\partial_{C_1}\overrightarrow\partial_{C_2}\right)e^{W}\;\;.\nonumber\\
\label{hs.5}
\end{eqnarray}
In the above, summation over repeated indices is understood and, for clarity
of the diagrams, the
Poisson decorations of the vertices have been omitted.

\end{itemize}

It is evident from these illustrations, as well from
Eqs.(\ref{zpoisson1})and (\ref{zpoisson2}) for  $H_n$,
that all the graphs for bi-differential
operators resulting from the Hausdorff series are non-loop graphs such that a graph with $n$-vertices is  formed by concatenation of a single wedge to a non-loop $n-1$-graph, allowing for the feet of the wedge to land either both on aerial vertices or one of them on a ground vertex and so that all the aerial
vertices of the resulting $n$-graph have one leg from an ordered wedge landing on them, with
the exception of the outermost aerial vertex.
The category of these graphs is refered to in the
literature as L-graphs.\\

Symbolically the bi-differential operator (\ref{2.9f}) can therefore
be expressed as
\begin{eqnarray}
&&\hat D\left (Z^{\ast}, f, ({\overleftarrow{\partial}}_{B_{1}},
{\overleftarrow{\partial}}_{B_{2}}\dots),
({\overrightarrow{\partial}}_{C_{1}},
{\overrightarrow{\partial}}_{C_{2}}\dots)\right )= \nonumber\\
&&\exp\left[\frac{1}{2} {\includegraphics[width=.4cm]{nb.eps}} +
\frac{1}{12}\left( {\includegraphics[width=.6cm]{nnbb1.eps}}+
{\includegraphics[width=.6cm]{bbnn.eps}}\right)
+\frac{1}{48}\left( {\includegraphics[width=.6cm]{nnnbbb.eps}}+
{\includegraphics[width=.6cm]{bbbnnn.eps}}\right) + \dots \right].\nonumber\\
\label{kontope}
\end{eqnarray}
\section{Relation to Kontsevich's Quantization}
\label{Kont}

In order to relate our preceding results with Kontsevich's
$\star-$product for deformation quantization let us begin by
reviewing briefly, both for self-containment and to fix notation,
the essentials of that construction
\cite{Kont:97,Dito}.

Let $G_n,\; n\geq 0$ denote the class of admissible graphs, {\it
i.e.} the class of oriented, labeled graphs $\Gamma \in G_n$ which
diagrammatically are associated to all bi-differential operators
that can be constructed from $n$ wedges. Let $V_{\Gamma}$ be the
finite set whose elements are vertices of $\Gamma$, and $E_\Gamma$
the finite set whose elements are the edges of $\Gamma$.

\begin{Def}
An oriented graph $\Gamma$ is a pair $(V_\Gamma, E_\Gamma)$ such
that $E_\Gamma \subseteq V_\Gamma \times V_\Gamma$.

\end{Def}

For $e=(v_1,v_2) \in E_\Gamma$ we say that the edge $e$ starts at
the vertex $v_1$ and ends at the vertex $v_2$.

A labeled graph $\Gamma$ belongs to $G_n$ if:
\begin{itemize}
  \item{1)} $\Gamma$ has $n+2$ vertices and $2n$ edges.
  \item{2)} The set $V_\Gamma$ is $\{1,2,\dots,n \} \cup {L,R}$,
  where $L,R$ are just two symbols meaning Left and Right and
  label the two ground vertices. These will correspond to the
  exponential functions $\exp(V)$ and $\exp(W)$ respectively,
  introduced in the previous section.
  \item {3)} There are two edges starting at every aerial vertex
  $k\in \{1,2, \dots, n\}$. These are ordered and labeled by
  $e^1_k, \ e^2_k$. In our notation of indexing with rooted trees,
  we have the correspondence $e^1_k\leftrightarrow B_k, \
  e^2_k\leftrightarrow C_k$.
  \item{4)} For any $v\in V_\Gamma, \ (v,v) \not\in E_\Gamma$, {\it
  i.e.} $\Gamma$ has no loop (an edge starting at some vertex and
  ending at the same vertex). However, graphs with loops (also
  called wheels) formed by an edge starting at vertex $i$ and
  ending at vertex $j$ and another edge starting at vertex $j$ and
  ending at vertex $i$ are allowed ("valid" loops).
  \item{5)} For $n\geq1$, the set $G_n$ is finite and has
  $(n(n+1))^n$ elements, and one element for $n=0$.
\end{itemize}

To each labeled graph $\Gamma\in G_n$ one can associate a
bi-differential operator
\begin{equation}\label{bgaal}
B_{\Gamma,\alpha}: A \times A \to A, \ \ A=C^\infty({\cal
U}), \ {\cal U}\ \hbox{an open domain of}\ \mathbb{R}^d
\end{equation}
given by the general formula
\begin{eqnarray}
  B_{\Gamma,\alpha} = \sum_{I:E_\Gamma \to \{1,\dots,d\}}
  \left(\prod_{e\in E_\Gamma, e=(*,L)}
  \overleftarrow{\partial}_{I(e)}\right)&&\left[\prod_{k=1}^n
  \left(\prod_{e\in E_\Gamma, e=(*,k)}\partial_{I(e)}\right)
  \alpha^{I(e_k^1)I(e_k^2)}\right]  \nonumber\\
  &&\times \left( \prod_{e\in E_\Gamma, e=(*,R)}
  \overrightarrow{\partial}_{I(e)}\right).\label{Bee}
\end{eqnarray}
Here $\alpha \in \Gamma({\cal U}, \bigwedge^2T_{\cal U})$ is a
bi-vector field (not necessarily a Poisson one), and the map
$I:E_\Gamma \to \{1,2,\dots,d\}$, replacing the labels $e_m^n$ by
independent indices, corresponds to the coloring of edges used in
the previous section, and is also useful in defining the coloring
for the vertices.

The next step in the Kontsevich construction consists in attaching
a weight $w_\Gamma\in \mathbb{R}$ to each graph $\Gamma\in
G_n$. The construction uses a special angular measure defined on
the Poincar\'e plane. The actual computation of the weights for
the most general graphs can become rather complicated and there
are questions of uniqueness in the results from different
calculation procedures \cite{Polyak}.

The associative Kontsevich $\star$-product is then defined as
\begin{equation}\label{kontp}
    f\star g := f\left( \sum_{n=0}^\infty \varepsilon^n \sum_{\Gamma\in G_n}
    w_\Gamma B_{\Gamma,\alpha}\right) g,
\end{equation}
where $\varepsilon$ is the deformation parameter, and $f, \ g \in
A$.

Our task is then to show that for a linear Poisson structure
determined by the structure constants of the Lie algebra ${\cal
L}$ introduced in Sec.2, Eqs.(\ref{2.8f}) and (\ref{kontp}) are the
same and that the Kontsevich operator is of the form
(\ref{kontope}), {\it i.e.} the exponential of a sum of prime
graphs (graphs that have no factors other than $\Gamma_0$ and
themselves under multiplication).
Thus composite graphs, resulting from multiplication of two prime
graphs by merging their respective left and right ground vertices
as in
\begin{equation}\label{paste}
\Gamma_A =\Ttbbnn{V}{W}{}{}{}{} \ \ \ \ \Gamma_B = \Alvwt{V}{W}
\Rightarrow \Gamma_{AB}=\Tgammab{V}{W},
\end{equation}
occur only through the expansion of the exponential.

To this end let us first set $\varepsilon =1$ in (\ref{kontp})
and, since the dual algebra ${\cal L}^*$ has infinite generators,
we also extend $\mathbb{R}^d \to \mathbb{C} -0$ and $A \to {\mathcal A}$ in (\ref{bgaal}). This poses no conceptual
problem since the functions $f,g$ occuring in (\ref{kontp}) will now be exponentials where the exponents, for any tree with a finite number of vertices, will have only a finite number of terms contributing to the calculation.
 Next note that
since we are dealing with a linear Poisson structure determined by
the structure constants $f^{A_l}_{B_iC_j}$, there are no diagrams
in the Kontsevich formula with two legs landing on any aerial
vertex. Also note that valid loops (wheels) would involve cyclic products
of the form
\begin{equation}\label{efes}
f^{C_n}_{B_1C_1} f^{C_1}_{B_2C_2} \dots f^{C_{n-1}}_{B_nC_n},
\end{equation}
as may be seen from the generic loop
\begin{equation}\label{loop1}
\Tloop{\partial_{C_{n-1}}}{n}{\partial_{C_{n}}}{f^{A_1}_{B_1C_1}Z^*_{A_1}}
{\partial_{C_{1}}}{f^{A_2}_{B_2C_2}Z^*_{A_2}}{k-1 \ \ \
\partial_{C_{k-1}}}{f^{A_k}_{B_kC_k} Z^*_{A_k}}{\partial_{C_{k}}}
\end{equation}

But by virtue of (\ref{1.4}) one gets that
\begin{equation}\label{ces}
    C_n > C_1 > C_2 > \dots  C_{n-1} >  C_n.
\end{equation}
This is impossible (since it would imply that a tree is of a higher degree than itself), and in consequence there are no valid loop
diagrams in the Kontsevich formula for the Poisson structure induced by $\mathfrak L$.\\

 Thus, in the linear setting and our specific structure constants,
 the only admissible prime diagrams for the Kontsevich $\star$-product are non-loop L-graphs, as was the case for the BCH $\star_1$-product.
So, in order to complete our argument that these two products are equivalent, irrespective of any particular expression for the Hausdorff series, we only need to show that the weights for their corresponding L-graphs are the same.
The essentials for the proof are contained in Theorem 5.0.2 in \cite{Kathotia} ({\it cf.} also \cite{Arnal, Polyak, Gutt}), which makes use of Properties (A 1) and (A 2) discussed in the previous section.\\

 Indeed, for two $\star$-products on ${\mathfrak L}^{*}$ to be equivalent all that is needed is:\\

\begin{itemize}
\item a) To show that property (A 1) is satisfied by both products, and\\
\item b) To show that the two products are equal for products of the form $(V)^{n}\star W$.\\

\end{itemize}
The proof that $p_m \star q_n = p_m  q_n + \text{terms of degree}
< m+n$ for the BCH product $\star_1$ that we considered in Sec. 4
was already given there. To show it for the Kontsevich product
(\ref{kontp}) with the bi-differential (\ref{Bee}) specialized to
the linear setting is just as easy, as it only involves counting
the legs landing on aerial vertices  (which removes an equal
number of powers of the $Z_{A}^{*}$'s from the vertex decorations)
and the number of remaining legs landing on the two ground
vertices (which reduces by an equal number the added degree $m +n$
of the $p_m$ and $q_n$ polynomials). For each type of the
admissible diagrams the total degree is
always lowered.\\

We have only left to show that for products of the form $(V)^{n}\star W$ the
coefficients of the bi-differentials in the exponential of (\ref{kontope}), taking
into account our choice of Poisson structure,
 are the same as the corresponding ones
in the summand of the Kontsevich operator:
\begin{equation}
 \exp\left[ \sum_{n=1}^{\infty}\sum_{\Gamma \in G_{n,L}} w_{\Gamma}
 B_{\Gamma,\alpha} \right], \text {where}\;  \; G_{n,L}\subset G_n \;\;\text
 {is the subset of prime L-graphs}\label{op}.
\end{equation}
To this end, note that the diagrams associated with products of
the form $(V)^{n}\star W$ are formed by succesive concatenations
of the
$\hbox{ad}_V\left(\Alvw{V}{}\right)$ graphs to the basic wedge, by
attaching the $V$ vertex of $\hbox{ad}_V $
to the
$V$ vertex of $\Alvw{V}{W}$ and landing the other foot of
$\hbox{ad}_V$ on any of the free aerial vertices. As may
be infered from (\ref{hs.1}), (\ref{hs.2}) and  (\ref{kontope}), the weights of these graphs for the BCH-quantization are given ({\it cf}  Eq.(\ref{zpoisson2})) by $2^{n}\frac{{\tilde B}_n}{n!}$, where ${\tilde B}_n =(-1)^n B_n $
and $B_n$ are the Bernoulli numbers ($B_n =0$ for $n>1$ odd.  Alternatively the Kontsevich
weights for these graphs, calculated with the angle measures described
in \cite{Kathotia}, are given by $ w_\Gamma = \frac{{\tilde B}_n}{(n!)^2}$. But  the diagrams that we used in
(\ref{kontope}) to express $\hat D$ symbolically, are actually
representatives of a class of diagrams of the same topological
type which differ by the labeling of the vertices and edges.
In fact, the class of topological type $[\Gamma_n]$
consists of $n! 2^n$ distinguishable graphs (we can label the $n$
vertices in $n!$ different ways, and there are two choices for the
ordering of the two edges that emanate from each vertex). On the
other hand the $\Gamma$'s that appear in the Kontsevich
bi-differential are individual graphs and the summation is done
over all individuals. But all the graphs belonging to a given
topological class lead to the same bi-differential operator, both
because the labeling of the vertices is irrelevant to the process
of assigning bi-differential operators and because the change in
sign resulting from the flipping of two edges from a given vertex
(due to the antisymmetry of the $\alpha$'s) is compensated by the
change in sign of the weight factor of the graph \cite{Kathotia}.
We can therefore replace the sum over all $\Gamma$'s in the
exponential of (\ref{op}) by the sum of representatives of the
corresponding topological class, with the {\it proviso} of
multiplying each term in the chain by $n! 2^n, \
n=1,2,\dots,\infty$. Consequently the weights of the diagrams originating from products of the form
$(V)^{n}\star W$, both in
(\ref{kontope}) and in (\ref{op}) are the same.
It then follows from the universality of the
BCH and Kontsevich quantizations that the weights in both
quantizations are the same for all the prime L-graphs. That is,
for renormalized pQFT,

\begin{equation}\label{equivk}
    \exp\left[ \sum_{n=1}^{\infty}\sum_{\Gamma \in G_{n,L}}
    w_{\Gamma} B_{\Gamma,\alpha} \right] = \hat D \left(Z^*, f,
    (\overleftarrow{\partial}_{B_1},\overleftarrow{\partial}_{B_2},
    \dots),(\overrightarrow{\partial}_{C_1},\overrightarrow{\partial}_{C_2},
    \dots)\right),\;\; G_{n,L}\subset G_{n},
\end{equation}
thus proving our contention. \\

Moreover, using (\ref{2.8f}) we have that
\begin{equation}
\exp\left((\psi^A)_R Z_A^*\right)= \exp\left(S_R
(\psi^B)Z_B^*\right) \star \exp\left((\psi^C) Z_c^*\right),\label{ren}
\end{equation}
where the Kontsevich $\star-$product is given by the operator
$\exp\left[ \sum_{n=1}^{\infty}\sum_{\Gamma \in G_{n,L}}
  \omega_{\Gamma} B_{\Gamma,\alpha} \right ]$.\\
From (\ref{ren}) we can immediately infer that the renormalized
normal coordinate $(\psi^A)_R $, corresponding to a given rooted
tree labeled by the index $A$, is the coefficient of the
coordinate $Z_A^*$ which appears in the exponential after the
$\star$-product on the right has been evaluated. Also, because of the
commutativity of the Hopf algebras, and the associativity of their
coproduct and that of the twisted antipode in the mass independent
renormalization scheme \cite{Kre:02}, the renormalized
$(\phi^{A})_R$'s ({\it cf.} (\ref{2.3})) follow directly from
(\ref{1.13}). Thus
\begin{equation}
 (\phi^{A})_R =  (\psi^{A})_R + \sum_{m=2}^{\infty} \frac{1}{m!}n^{A}_{B_1 \dots B_m} (\psi^{B_1})_R \dots (\psi^{B_m})_R .\label{50}\\
\end{equation}

\section{The Birkhoff Algebraic Decomposition and the Kontsevich Deformation}

Our formulation of the Hopf algebra of renormalization in terms of normal coordinates
together with Eq.(\ref{2.7f}) for the renormalized representative of a rooted tree,
provide an immediate relation with the Birkhoff algebraic decomposition and, in turn,
a relation of the latter with the Kontsevich deformation.\\

In fact, as we have already seen from Eq.(\ref{2.5}) in Sec.3, the twisted antipode
projects the regularized normal coordinates $\psi^{A}\in {\mathcal H}_R$ onto
${\mathcal A}_{-}$. Therefore,
defining

\begin{eqnarray}
&\phi_{-}&:=e_{\ast}^{S_{\mathcal R}(\psi^{A})\otimes Z_A},\nonumber\\
&\phi&:=e_{\ast}^{\psi^{A}\otimes Z_A},\label{birk}\\
&\phi_{+}&:=e_{\ast}^{(\psi^{A})_R \otimes Z_A},\nonumber
\end{eqnarray}
and substituting into (\ref{2.7f}) immediately yields the Birkhoff factorization:
\begin{equation}
 \phi_{+}=\phi_{-}\ast\phi.\label{algbirk}
\end{equation}
Note, in particular, that
\begin{eqnarray}
\langle\phi_{-},({\text id}\otimes \psi^{A})\rangle&=&S_
{\mathcal R}(\psi^{A}),\nonumber\\
\langle\phi_{+},({\text id}\otimes \psi^{A})\rangle&=&(\psi^{A})_{R}.
\label{anti}
\end{eqnarray}

As a parenthetical remark, note also that we can relate the above
results with those in \cite{Kre:04} by writing $Z:=\psi^B\otimes
Z_B$ and $ S_{\mathcal R}(\psi^B)\otimes Z_B \equiv -{\mathcal
R}(\chi(Z)) \in{\mathcal A}_{-}\otimes {\mathfrak L}$, where
\begin{equation}
\chi(Z)=Z+\sum_{k=1}^{\infty} \chi_{Z}^{(k)}.\label{51}
\end{equation}
The $\chi_{Z}^{(k)}$'s in the above series are derived by iteration on the equation:
\begin{equation}
\chi^{(k)}_Z = \sum_{i=1}^{k} c_i K^{(i)}(-{\mathcal
R}(\chi^{(k-i)}_Z ), Z), \;\;\;\;\;\;\;\chi_{Z}^{(0)}\equiv
Z,\label{twist}
\end{equation}
where $K^{(k)}(-{\mathcal R}(Z), Z)$ are the nested multicommutators of
depth $k\in {\mathbb N}$ in the Hausdorff series, and the $c_k$ their corresponding
coefficients. \\
Thus calculating explicitly up to a depth 3 we have

\begin{eqnarray}
\chi_{Z}^{(1)}&=& -\frac{1}{2}[{\mathcal R}(Z), Z], \nonumber\\
\chi_{Z}^{(2)}&=&\frac{1}{4}[{\mathcal R}([{\mathcal R}(Z), Z]), Z] + \frac{1}{12}\left ( [{\mathcal R}(Z), [{\mathcal R}(Z), Z]] + [Z, [{\mathcal R}(Z), Z]]\right )\nonumber\\
\chi_{Z}^{(3)}&=& -\frac{1}{8}[{\mathcal R}([{\mathcal R}([{\mathcal R}(Z), Z]), Z]), Z]- \frac{1}{24}[{\mathcal R}(K^{(2)}(-{\mathcal R}(Z), Z)), Z]\nonumber\\
&+& \frac{1}{24}K^{(2)}({\mathcal R}([{\mathcal R}, Z], Z)+\frac{1}{48}K^{(3)}(-{\mathcal R}(Z), Z), \label{54}
\end{eqnarray}
and
\begin{eqnarray}
-{\mathcal R}(\chi(Z))&=&
- R \{\psi^{A}-\frac{1}{2}f^{A}_{B_{1}C_{1}} R(\psi^{B_1})\psi^{C_1}+ \frac{1}{4}f^{A}_{A_{1}C_{2}} f^{A_1}_{B_1 C_1} R(R(\psi^{B_1}) \psi^{C_1})\psi^{C_2}\nonumber\\
&+&\frac{1}{12}\left (f^{A}_{B_1 A_1} f^{A_1}_{B_2 C_1} R(\psi^{B_1})R(\psi^{B_2})\psi^{C_1}+
f^{A}_{C_1 A_1} f^{A_1}_{B_1 C_2} \psi^{C_1} R(\psi^{B_1}) \psi^{C_2}\right )\dots  \}\otimes Z_{A}.\nonumber\\\label{55}
\end{eqnarray}
The last expression is of course precisely the same as the one we would
obtain for $S_{\mathcal R}(\psi^{A})\otimes Z_A $ by making use of (\ref{2.5}).\\

Also, making  use of the BCH formula for the Hausdorff series
together with (\ref{51}) and (\ref{twist}), one has the following alternate
expression for $\phi_{+}$ ({\it cf.} \cite{Kre:04}):

\begin{equation}
\phi_{+} = e_{\ast}^{(1-{\mathcal R})(\chi(Z))}.\label{57}\\
\end{equation}

Let us now return to (\ref{algbirk}) with $\phi$, $\phi_{-}$ and $\phi_{+}$ given
by (\ref{birk}) (or their equivalent expressions in terms of $Z$ and $\chi(Z)$ given
above). Comparing with the first equality in (\ref{2.7f}), they are clearly
the same. Hence we can make use of (\ref{2.8f}), and (\ref{equivk}) to conclude
that
\begin{eqnarray}
\langle\phi_{+}, {\text id}\otimes T^{A}\rangle\otimes Z_{A}&=& \langle\phi_{-} \ast \phi, {\text id}\otimes T^{A}\rangle\otimes Z_A \nonumber\\
&\Leftrightarrow& e^{-{\mathcal R}(\chi(Z^{*})} \exp\left[ \sum_{n=1}^{\infty}\sum_{\Gamma \in G_{n,L}}
    w_{\Gamma} B_{\Gamma,\alpha} \right]e^{Z^{*}}, \label{58}
\end{eqnarray}
where on the right side we have used the notation $Z^{*}\equiv\psi^{A}Z^{*}_{A};\;\;\;Z^{*}_{A}\in {\mathfrak L}^{*}$.\\

That is, renormalization of pQFT encoded in the Birkhoff algebraic
decomposition can be viewed as a deformation, via the Kontsevich
product $\phi_{-} \star \phi$, of the pointwise multiplication of
the exponential functions $\phi_{-}$ and $\phi$ (expressed in
terms of coordinates of the dual Lie algebra ${\mathfrak L}^{*}$,
as it was done in Sec.4) in the direction of the linear Poisson
bracket.

\section{Conclusions}
In a previous paper \cite{Chry}, normal coordinates were introduced
in the context of the Hopf algebra formalism of renormalization for the
purpose of studying primitive elements of this algebra. It was shown there
that the use of normal coordinates lead naturally to the concept of k-primitiveness,
associated with the lower central series of the dual Lie algebra. For ladder trees
with the same decoration on all vertices, it was also shown that normal coordinates
provided remarkable simplifications in the renormalization process. Because of their
specific relation to rooted trees ({\it cf.} Eq.(\ref{1.15})) it is natural to expect
that the ensuing simplified pole structure of the regularized normal coordinates,
relative to that for rooted trees, will persist even for
branched trees with multiple decorations on the vertices.\\
In the present paper we have shown that by further using the properties of the Hopf
algebra of normal coordinates, a natural relation can be established between the
twisted antipode axiom of renormalization (the Forest Formula for pQFT) and the BCH
quantization formula. We also showed the equivalence of this $\star$-product to the
universal Kontsevich $\star$-product for deformation quantization, and that of the
latter to the Birkhoff algebraic factorization. Last, but not least, we showed that
for pQFT the Kontsevich product is of the form of an exponential of a sum of weighted
prime L-graphs.\\

 There are other studies in the direction of establishing a connection
between star products, Hopf algebras, and quantum groups in field
theory \cite{Pinter,Hir,Hir1,Brouder,Brouder1}. In particular in
\cite{Hir1}, the time-ordered product in field theory is related
to the Weyl transform of a Drinfeld twisted product. It would be
interesting to try to relate our work with these studies and see
if from the combination of both approaches it is possible to
obtain renormalized time ordered products and a more direct
physical interpretation for the normal coordinates.

\section*{Acknowledgements} The authors acknowledge partial support from
CONACyT project G245427-E (M. R.) and DGAPA-UNAM grant IN104503-3
(J.D.V.).

\end{document}